\documentclass[a4paper]{article}
\usepackage[utf8]{inputenc}
\usepackage[colorinlistoftodos]{todonotes}
\usepackage{graphicx}

\usepackage{geometry}
\usepackage{cite}
\usepackage{titling}
\usepackage{wrapfig}
\usepackage{lipsum}

\usepackage[square,sort,comma,numbers]{natbib}
\title{\textbf{Energy Harvesting Powered Embedded Systems}}
\author{Yawen Wu, Zhenge Jia, Jingtong Hu\\
Department of Electrical and Computer Engineering, University of Pittsburgh}
\date{}
\begin{document}

\maketitle

\vspace{-55pt}
\section{Introduction}
\vspace{-8pt}
Historically, battery is the power source for mobile, embedded and remote system applications. However, the development of battery techniques does not follow the Moore's Law. The large physical size, limited  electric quantity and high-cost replacement process always restrict the performance of the application such as embedded systems, wireless sensors networks and lower-power electronics. Energy harvesting, a technique which enables the applications to scavenge energy from RF signal from TV towers, solar energy, piezoelectric driven by motion of people and thermal energy from the temperature difference, which could dramatically extend the operating lifetime of applications. Thus, energy harvesting is important for the sustainable operations of an application.





\begin{wrapfigure}{r}{10.5cm}
\centering
\vspace{-2pt}
\includegraphics[width=300pt]{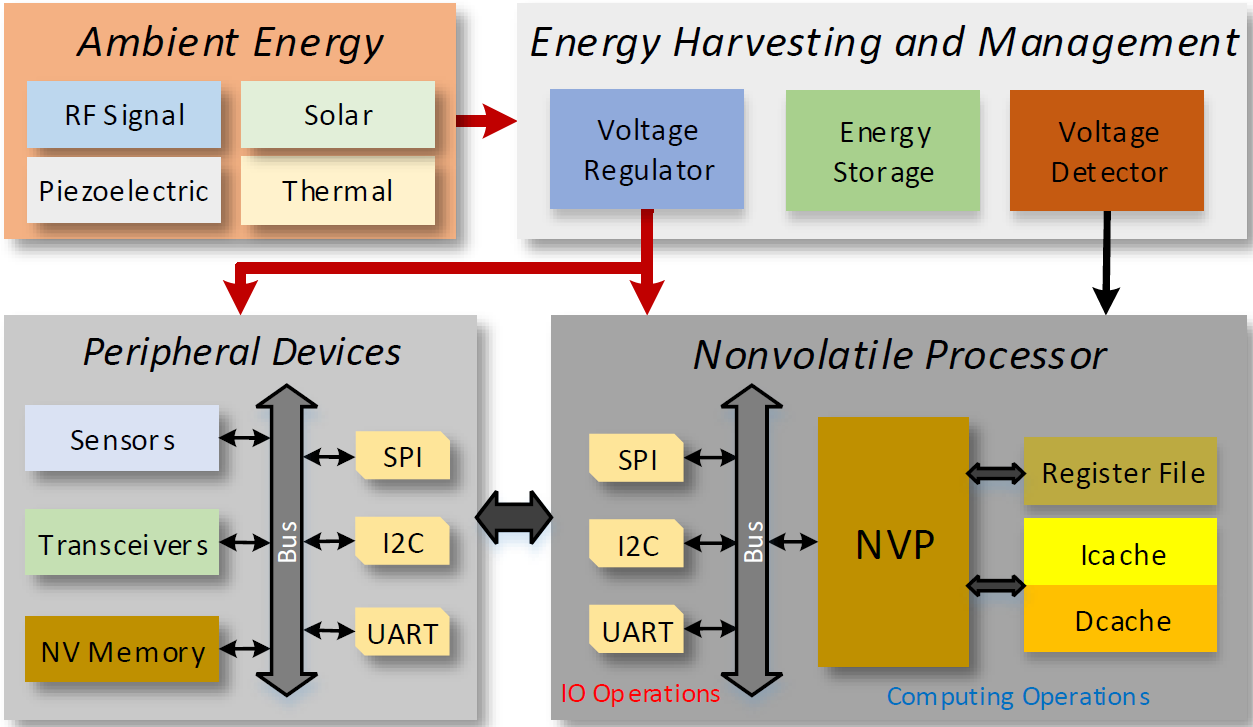}
\vspace{-20pt}
\caption{\small \sl Architecture of Energy Harvesting Powered Sensing System.\label{fig:arch}}  
\vspace{-10pt}
\end{wrapfigure}

\vspace{-10pt}
\noindent
\newline
Energy harvesting powered sensing systems are emerging as a key technology for Internet of Things (IoT) and Wireless Body Area Network (WBAN)~\cite{ku2016advances}. The power consumption of a typical sensor node in such systems ranges from several $uW$ to hundreds of $mW$, based on the task undertook by the node. Fig. \ref{fig:arch} is the typical architecture of an energy harvesting powered sensing system. Ambient energy can be harvested by a harvester, 
{which has high manufacturing complexity and low energy utilization.} 
{Fortunately, small on-chip inductors make energy harvesting circuits easier to implement and further improve the energy conversion efficiency~\cite{tida2014through, tida2014novel, tida2015efficacy, tida2014opportunistic}.
}The energy management module regulates the harvested to power processors and multiple peripherals. The processor, which is usually a Nonvolatile Processor (NVP)~\cite{xie2014non} equipped with Nonvolatile Memory (NVM), can budget the overall energy for each module and manage the behavior of the energy management module. One advantage of NVP compared with conventional processor is it can survive power shortage without losing their status, which fits well with the intermittent characteristic of the ambient power. The peripheral devices includes sensors, transceivers and NVM. Sensors are used to collect data, such as temperature, humidity and barometric pressure from the environment. Transceivers transmits the collected data to a remote node or router node, which forwards the data further to a server. NVM is used to maintain in-process data between power outages and store configurations such as local address for communication.

\vspace{-12pt}
\section{Related Works in Energy Harvesting}
\vspace{-8pt}
In this section, we will first introduce the energy harvesting applications, which harvest energy from ambient power sources and use the harvested power to complete operations. Then we will introduce several optimization techniques to energy harvesting based systems, which improve system stability and computing performance. 
\vspace{-10pt}
\subsection{Energy Harvesting Applications}
\vspace{-2pt}
The Ultra-Low Power Sensor Evaluation Kit (ULPSEK)~\cite{tobola2018self}-for evaluation of biomedical sensors and monitoring applications, is a wearable, multi-parameter health sensor powered by an efficient body heat harvester. ULPSEK could measure and process electrocardiogram, respiration, motion and body temperature. The key component of ULPSEK is the thermal harvester, which is placed at the forearm or the chest. The harvester consists of a heat sink, a thermal-electric module and a DC-DC converter circuit. 
As for mechanism energy, a Piezoelectric Energy Harvester (PEH) is proposed to be used in cantilever configuration and Structural Health Monitoring (SHM)~\cite{balguvhargreen}. It harvests energy from the vibration of bridge caused by the passing vehicles in the form of concrete vibration sensor for reinforced concrete structure. 
RF energy scavengers are the field  attracting a great deal of researchers' attention, especially in low-power wireless sensors networks. Researchers from the University of Tokyo propose a deign of low-cost sensor nodes harvesting energy from TV broadcast signal and storing excess power in capacitors~\cite{shigeta2013ambient}. 

\vspace{-10pt}
\subsection{Advances in Optimization and Design Process}
\vspace{-4pt}
There is an intrinsic drawback with energy harvesting application, that is they are intermittent. Since almost all traditional computer systems are designed to be with a stable power supply, they could not make much progress with the intermittent power sources. 
Thus, it is necessary to implement optimizations in different perspectives to enable the devices to complete assigned tasks with intermittent power sources and other constraints. 
\newline
\noindent
\textbf{Progress Accumulation:} To make progress, we have to accumulate the computing across intermittent power cycles. One of the key techniques to achieve the goal is checkpointing, which save the processor's states to a non-volatile memory before a power failure and resume from the stored state when power comes back on.
The domain of intermittent computing contains a significant amount of work, which address power loss during a devices operation and make a checkpointing~\cite{xie2014non}.
The adoption of NVP in energy harvesting devices is emerging as a decent solution to the intermittent computing. 
There are several compiler optimization techniques have been published and these works show a promising results for the utilization of NVP~\cite{xie2014non}. 
\noindent
\newline
\textbf{Power Management:} The energy buffer in energy harvesting devices is typically recharging batteries and capacitors, which all have limited capacities and so any energy generated when the energy buffer is full will be wasted. Thus, it is necessary to develop an efficient power management to achieve the maximization usage of the harvested power. 
There are several power management achieved by tuning different software and hardware parameters in the energy harvesting devices, such as duty cycles~\cite{vigorito2007adaptive}, sensing rate and transmitting power.

\vspace{-12pt}
\section{Challenges and Research Opportunities}
\vspace{-6pt}
Although the recent advances in energy harvesting powered embedded systems, there still exists challenges and opportunities for future researches.




\noindent
\textbf{Energy Harvesting Sensing Network with Multiple Nodes:} 
Although extensive researches have been carried on energy harvesting sensor node, there are still some challenges when adopting these techniques to the whole sensing network consisting of multiple nodes. 
When sensor nodes are dispersively deployed in the field, some far-away nodes cannot communicate with the router node directly due to signal dissipation. Low power routing algorithm
is desired to forward the information from the unreachable node back to the router. Multi-hop information
even energy relay and overall information and energy cooperation applicable to energy harvesting node communication are still challenging because of symmetry of energy consumption of transmitting and receiving data per time unit. But relatively longer time window for receiving in Time Division Multiple Address (TDMA) network leads to higher energy consumption than transmitting the same length packet.


\noindent
\textbf{Health Concerns:} It has been long recognized RF exposure can cause heating of materials including biological tissues. 
Some research~\cite{breckenkamp2009feasibility} shows that genes can be affected when RF power approximates the upper bound of international security levels. Although there are some studies on the  effects of RF exposure from mobile and cellular networks, health effects caused by dedicated RF power transmitter which emits much higher power than cell phones are still required. Moreover, one way to mitigate health concerns is to decrease the transmitting power of a dedicated power source. However, this evicts challenges to the RF harvesting sensor node on how to leverage the limited power to finish predefined task efficiently.



\vspace{-12pt}
\bibliographystyle{unsrt}
\bibliography{ref.bib}

\end{document}